\documentclass[preprintnumbers, prd, twocolumn, showpacs, floatfix,
preprintnumbers,  superscriptaddress, nofootinbib]{revtex4}
\usepackage[dvips]{graphicx}
\usepackage{epsf}
\usepackage{amsmath}
\usepackage{amssymb}
\usepackage{graphicx}
\usepackage{dcolumn}
\usepackage{bm}
\usepackage{color}
\usepackage{epsfig}
\usepackage{amsfonts}
\usepackage{ulem} 

\def\be{\begin{equation}}
\def\ee{\end{equation}}
\def\ba{\begin{eqnarray}}
\def\ea{\end{eqnarray}}

\def\be{\begin{equation}}
\def\ee{\end{equation}}
\def\bea{\begin{eqnarray}}
\def\eea{\end{eqnarray}}
\def\fft#1#2{{\frac{#1}{#2}}}

\begin{document}
\title{Constraining Ho\v{r}ava-Lifshitz gravity from neutrino speed
experiments}

\author{Emmanuel N. Saridakis}
\email{Emmanuel_Saridakis@baylor.edu}
 \affiliation{CASPER,
Physics Department, Baylor University, Waco, TX  76798-7310, USA}

\begin{abstract}
We constrain Ho\v{r}ava-Lifshitz gravity using the results of OPERA and
ICARUS neutrino speed experiments, which show that neutrinos are luminal
particles,  examining the fermion propagation in the earth's gravitational
field. In particular, investigating the Dirac equation in the spherical
solutions of the theory, we find that the neutrinos feel an effective
metric with respect to which they might propagate superluminally.
 Therefore, demanding not to have superluminal or subluminal motion we
constrain the parameters of the theory. Although  the excluded parameter
regions are very narrow, we find that the detailed balance case lies in
the excluded region.
\end{abstract}


\maketitle

\section{Introduction}

In September 2011 the OPERA collaboration announced the astonished
result that muon-neutrinos created in CERN CNGS beam in Geneva were
detected in Sasso Laboratory in central Italy, faster than the time light
would need to cover the same distance in vacuum  \cite{:2011zb}. In
particular, it was reported that  $\nu_\mu$ neutrinos arrive earlier than
expected from luminal speed by a time interval 
\begin{eqnarray}\delta t = 57.8\pm
7.8(stat.)^{+8.3}_{-5.9} (syst.)\ \text{ns},
\end{eqnarray}
corresponding to  a superluminal propagation of an
amount 
\begin{eqnarray}
 (v-c)/c
=(2.48\pm0.28({\text{stat}})\pm0.30({\text{sys}}))\times10^{-5},
\end{eqnarray}
where $v$ is the nuetrino velocity and $c$ the light speed, 
a result that is in agreement with earlier 1$\sigma$ MINOS announcements
\cite{Adamson:2007zzb}. However, in February 2012, the OPERA
collaboration announced that the ``measured'' superluminality was a result
of a loose fibre optic cable. Indeed, on March 2012 the ICARUS
collaboration, also using the CNGS neutrino beam, measured that the time of
flight difference between the speed of light and the arriving neutrinos
was \cite{Antonello:2012hg}
\begin{eqnarray}\delta t =0.3\pm
4.9(stat.)^{+9.0}_{-9.0} (syst.)\ \text{ns},
\end{eqnarray}
 which is compatible with the simultaneous
arrival of all events at a luminal speed.

Although the possibility of systematic errors was the reasonable
explanation straightaway from the beginning, the OPERA
announcement attracted the
interest of theorists, who tried to explain it following many different
paths: in \cite{Cacciapaglia:2011ax,Mann:2011rd,Lingli:2011yn} proposing
simple models of Lorentz violation by hand, in
\cite{Li:2011ue} imposing a mass-dependent Lorentz violation,
in \cite{AmelinoCamelia:2011dx} using models of energy-dependent
velocities, in \cite{Klinkhamer:2011mf} with a Fermi-point splitting, in
\cite{Anacleto:2011bv} using noncommutativity, in \cite{Wang:2011zk} using
monopoles, in
\cite{Giudice:2011mm} imposing Lorentz violations through
light sterile
neutrinos \cite{Pas:2005rb}, in
\cite{Gubser:2011mp,Nicolaidis:2011eq,Hannestad:2011bj}
assuming that neutrinos can propagate
in extra dimensions, in \cite{Pfeifer:2011ve} using Finsler spacetimes,  in
\cite{Konoplya:2011ag} using G\"{o}del-like rotating universe, in
\cite{Magueijo:2011xy} using tachyonic mixed neutrinos (although in
\cite{Drago:2011ua} it was shown that superluminality cannot be explained
using tachyonic or Coleman-Glashow models), in
\cite{Wang6930} using a domain wall, in \cite{Ciuffoli:2011ji} using a
neutrino-scalar coupling, in \cite{Matone:2011jd} using the quantum
Hamilton-Jacobi equation, and in \cite{Gardner:2011hg} through a truer
measurement of Einstein’s limiting speed. Additionally, in
\cite{Alexandre:2011kr} and
\cite{Alexandre:2011bu,Klinkhamer:2011iz,Li:2011zm} a
Lifshitz-type, Lorentz-violated fermion model
\cite{Colladay:1998fq,Kostelecky:2003cr} was assumed, while in
\cite{Dvali:2011mn,Iorio:2011ay,Franklin:2011ws} and \cite{Kehagias:2011cb}
the authors proposed the
interesting idea of a local effective neutrino supelumination, without
Lorentz voilation, due to a coupling with a new spin-2 field or a scalar
respectively. Finally, we have to mention that straightaway from the
beginning  there were works claiming
that the superluminal interpretation was not correct: in
\cite{Ciborowski:2011ka,Bi:2011nd,Cohen:2011hx} it was argued that
the superluminal
neutrinos would decay through a number of channels, while in 
 \cite{Contaldi:2011zm} the author put into question the
convention for
synchronization of clocks in non-inertial frames and in
\cite{Broda:2011du,Torrealba:2011ww,Winter:2011zf} various other possible
systematic errors were discussed.

In the present work, inspired by the works claiming that neutrino
supeluminality is a local effect around earth
\cite{Dvali:2011mn,Iorio:2011ay,Franklin:2011ws,Kehagias:2011cb},
we investigate the neutrino propagation in the effective background
metric of earth's gravitational field,
in the context of Ho\v{r}ava-Lifshitz gravity. Since supeluminality may be
the case in a small region of the parameters of the theory, we use this
result in order to constrain the parameters of Ho\v{r}ava-Lifshitz in
order to be consistent with OPERA and ICARUS collaboration non-superluminal
results.

\section{Spherical solutions in Ho\v{r}ava-Lifshitz gravity}

Let us briefly review the spherical solutions of simple Ho\v{r}ava-Lifshitz
gravity.  The dynamical variables are the lapse and shift functions, $N$
and $N_i$ respectively, and the spatial metric $g_{ij}$ (roman letters
indicate spatial indices). In terms of these fields the full
metric is written as $ds^2 = - N^2 dt^2 + g_{ij} (dx^i + N^i dt ) ( dx^j +
N^j dt )$, and the (anisotropic) scaling transformation of the coordinates
reads: $
 t \rightarrow l^3 t~~~{\rm and}\ \ x^i \rightarrow l x^i$.
As it is known, the action of the theory can be decomposed as
\cite{hor3,hor2,Saridakis:2009bv,Cai:2009in}
\begin{eqnarray}
 S &=&  \int dt d^3x\sqrt{g} N \left({\cal L}_0 + {\cal
L}_1\right)
\label{Stot}\\
 {\cal{L}}_0 
&=&
\frac{2}{\kappa^2}
(K_{ij}K^{ij} - \lambda K^2)+\frac{\kappa^2\mu^2(\Lambda R
  -3\Lambda^2)}{8(3\lambda-1)}\label{L0}\\
 {\cal L}_1 &=&
 \frac{\kappa^2}{2 w^4} C_{ij}C^{ij}
 -\frac{\kappa^2 \mu}{2 w^2}
\frac{\epsilon^{ijk}}{\sqrt{g}} R_{il} \nabla_j R^l_k \nonumber\\
&\ &\ \ \ \ \ \ \ \ \  +
\frac{\kappa^2 \mu^2}{8} R_{ij} R^{ij}
-    \frac{\kappa^2 \mu^2(1
- 4 \lambda)}{32(3 \lambda-1)} R^2,
\label{L1}
\end{eqnarray}
where $ K_{ij} =   \left( {\dot{g_{ij}}} - \nabla_i N_j - \nabla_j
N_i \right)/2N $
 is the extrinsic curvature and
$ C^{ij}  =  \epsilon^{ikl} \nabla_k \bigl( R^j_{\ l} -  R
\delta^j_{\ l}/4 \bigr)/\sqrt{g} $ the Cotton tensor, and the
covariant derivatives are defined with respect to the spatial
metric $g_{ij}$. $\epsilon^{ijk}$ is the totally antisymmetric
unit tensor, $\kappa$, $w$, $\mu$ and $\Lambda$ are constants (we have
already performed the usual analytic continuation of the parameters $\mu$
and $w$ and thus $\Lambda$ is positive), and $\lambda$ is the
dimensionless constant that determines the flow between th IR and UV. 
We mention that in writing the above action splitting, and with
the particular coefficients, we have imposed the detailed balance condition
\cite{hor3}, which allows for a quantum inheritance principle \cite{hor2},
since the $(D+1)$-dimensional theory acquires the renormalization
properties of the $D$-dimensional one. Finally, it is straightforward to
see that for the light speed, the gravitational Newton's
constant\footnote{Note that in theories with Lorentz invariance breaking
the  ``gravitational'' Newton's constant, that is the one that is read from
the action,  does not coincide with the ``cosmological'' Newton's
constant, that is the one that is read from the Friedmann equations
\cite{Carroll:2004ai}, but this is irrelevant for the purposes of this work
where we focus on non-cosmological scales.} and the effective cosmological
constant we obtain:
\begin{eqnarray}
\label{physicalvalues}
 c=\fft{\kappa^2\mu}{4} \sqrt{\fft{\Lambda}{3\lambda-1}}\,,\ 
G=\fft{\kappa^2}{32\pi\,c}\,,\
\Lambda_{eff}=\frac{3\kappa^2\mu^2\Lambda^2}{16(3\lambda-1)}.
\end{eqnarray}
As one observes, the light speed flows too, however one can still set it to
$1$, and consider photons to propagate with this speed always, which will
be the reference speed in  Ho\v{r}ava-Lifshitz gravity.

Under different assumptions there are many spherical solutions in the
gravitational scenario at hand
\cite{Lu:2009em,Cai:2009pe,Kehagias:2009is,Kiritsis:2009rx}, which
extract the extra terms comparing to General Relativity. For the purpose of
this work we desire to remain in a general but still simple level. Thus, we
should go beyond the detailed balance condition, which proves to lead to
theoretical and observational problems
\cite{Charmousis:2009tc,Bogdanos:2009uj,Dutta:2009jn,Dutta:2010jh}, but
still keeping the structure of
the theory simple. Therefore, it is adequate to deform action (\ref{Stot})
as \cite{Lu:2009em,Cai:2009pe}
 \begin{eqnarray} S=  \int dt d^3x\sqrt{g} N \left\{{\cal
L}_0 + (1-\epsilon^2){\cal
L}_1\right\},
\end{eqnarray}
with $\epsilon$ a  parameter.

Seeking for static, spherically symmetric solutions with the metric ansatz
 \begin{eqnarray}
\label{sphericalsol}
ds^2 = - N(r)^2\,dt^2 + \fft{dr^2}{f(r)} + r^2 (d\theta^2
+\sin^2\theta d\phi^2),
\end{eqnarray}
and setting $\lambda=1$, as expected for earth scales, one obtains:
 \begin{eqnarray}
\label{solution}
N(r)^2&=&f(r) \nonumber\\
&=& 1 + \fft{\Lambda r^2}{1-\epsilon^2} -
\fft{\sqrt{\alpha^2(1-\epsilon^2) \sqrt{\Lambda}r + \epsilon^2
\Lambda^2 r^4}}{1 -\epsilon^2}.\ \ \ \
\end{eqnarray}
In this expression the integration constant $\alpha$ can be expressed in
terms of the total mass of the spherical object and the gravitational
Newton's constant \cite{Lu:2009em,Cai:2009pe}.

We mention here that the peculiar second term in (\ref{solution}), which
will play the central role in the following discussion, arises
in the majority of the corresponding solutions
\cite{Lu:2009em,Cai:2009pe,Kehagias:2009is,Kiritsis:2009rx}. For example,
if we take the Kehagias-Sfetsos (KS) model \cite{Kehagias:2009is} and
extend it to the minimal beyond-detailed-balance case (that is taking a
general coefficient of the Ricci scalar term in the action (\ref{Stot})) we
can obtain the KS spherical solution 
 \begin{eqnarray}
\label{solution22}
N_{KS}(r)^2=f_{KS}(r) = 1 + q r^2 - \sqrt{r(q^2 r^3+4q M G)},
\end{eqnarray}
where $q$ in now a free parameter, negative due to the analytic
continuation (in this work we have transformed the parameters $\mu$ and
$w$ of \cite{Kehagias:2009is} as $\mu\rightarrow i\mu$ and 
$w^2\rightarrow- i w^2$ \cite{Lu:2009em}), and $M$ is the total mass (note
that one could
alternatively obtain the above solution keeping the detailed-balance
version of KS model, but move slightly away from $\lambda=1$).

\section{Neutrinos motion in earth's gravitational field}

Let us now investigate the propagation of fermions, and in particular of
neutrinos, in earth's gravitational field. Considering a massive Dirac
field in a curved background $g_{\mu\nu}$, the equation of motion reads
\cite{DeWitt:1975ys}:
\begin{equation}\
\label{diraceuation}
    \left[\gamma^{a} e_{a}^{\mu}\left(\partial_{\mu} + \Gamma_{\mu}\right)
+ \frac{m}{\hbar}\right] \Psi = 0,
\end{equation}
where $m$ is the fermion mass and $\hbar$ the Planck's constant. In this
relation $e_{a}^{\mu}$ is the inverse of the vierbein tetrad
field $e^{a}_{\mu}$, defined as $g_{\mu\nu} = \eta_{ab}e^{a}_{\mu}e^b_\nu$
with $\eta_{ab} = \text{diag} (-1, 1, 1, 1)$, $\gamma^{a}$ are the Dirac
matrices (taken in the standard representation \cite{DeWitt:1975ys}) and 
$\Gamma_{\mu}$ is the spin connection given by
\begin{equation}
\label{Gamma}
    \Gamma_{\mu} = \frac{1}{8}\left[\gamma^{a},
\gamma^{b}\right]e_{a}^{\nu}e_{b\nu ; \mu},
\end{equation}
where the covariant derivative of $e_{b\nu}$ is as usual $ e_{b\nu; \mu} =
 \partial_{\mu} e_{b\nu} - \Gamma_{\mu\nu}^{a}e_{ba}$.
 
Let us investigate the Dirac equation in the earth's background,
considering that its gravitational field is given by (\ref{sphericalsol}),
that is in a vierbein reading as
\begin{equation}
\label{HLvierb}
e_{a}^{\mu} = \text{diag} \left(\frac{1}{\sqrt{f}},  \sqrt{f},
\frac{1}{r},  \frac{1}{r\sin\theta}\right),
\end{equation}
with $f(r)$ given by (\ref{solution}) or (\ref{solution22}).
Neglecting for simplicity the spin connection $\Gamma_{\mu}$, which proves
to be vary small, the Dirac equation (\ref{diraceuation}) under the
geometry (\ref{HLvierb}) reads:
{\small{
\begin{equation}
\label{Dirac2}
    \left(\frac{\gamma^0}{\sqrt{f(r)}}\partial_t + \sqrt{f(r)} \gamma^1
\partial_r +
    \frac{\gamma^2}{r} \partial_\theta + \frac{\gamma^3}{r\sin\theta}
\partial_\phi +
 \frac{m}{\hbar} \right) \Psi = 0.
\end{equation}}}
From this relation one can immediately see that the neutrinos feel an
effective metric, and that their velocity is simply
\begin{equation}
v(r)= f(r),
\end{equation}
and in particular if they propagate
in an approximately
constant $r$, equal for instance with the earth's radius $r=R_\oplus$,
their speed will be $v= f(R_\oplus)$=const..

 Additionally, we can verify
this result by approximately solving the Dirac equation
(\ref{Dirac2}) under certain assumptions. In particular, in the standard
Dirac matrices representation the fermion wave function is written as
\begin{eqnarray}
\label{Psirep}
 \Psi (t,r,\theta,\phi) 
&=& \begin{pmatrix}
 A (t,r,\theta,\phi) \\ 0 \\ B (t,r,\theta,\phi) \\
 0 \end{pmatrix} \exp \left[\frac{i}{\hbar}
I(t,r,\theta,\phi)\right].\ \ 
\end{eqnarray}
Without loss of generality, and in order to avoid difficulties of solving
Dirac equation in spherical coordinates, and using the spherical symmetry,
we can assume that $A$ and $B$ are constants, while $I(t,r,\theta,\phi)=-
\omega t + p(r) r + \Theta (\theta, \phi)$ \cite{Liu:2011zzt}, with $p(r)$
the neutrino momentum. In such a case the two relevant equations read 
\begin{eqnarray}
  -\frac{A}{\sqrt{f(r)}} \omega + B\sqrt{f(r)} p(r) + m
A &=& 0
 \nonumber\\
  \frac{B}{\sqrt{f(r)}} \omega - A\sqrt{f(r)} p(r)  + m B &=& 0,
\end{eqnarray}
and thus the solution condition (the determinant of $A$, $B$
coefficients to be zero)
leads to the dispersion relation 
\begin{equation}\label{omega}
 \omega^2=f(r)^2 p(r)^2+m^2f(r).
\end{equation}
In the massless case we can see that both the group velocity
$\partial\omega/\partial p$ and the phase velocity $\omega/p$ are equal to
$f$, that is $v(r)= f(r)$ .

In summary, we showed that the neutrino's velocity in the earth's
gravitational field in Ho\v{r}ava-Lifshitz gravity is equal to
$v=f(r)$, with $f(r)$ given by (\ref{solution}) or (\ref{solution22})
according to the specific solution subclass one uses. Observing the form of
$f(r)$ we can clearly see that $v(r)$ may becomes superluminal, that is
$v(r)>1$. Clearly this is not
the case in General Relativity spherical solutions, where the examination
of the Dirac equation, similarly to the above procedure, leads always to
$v<1$. 

Let us now come to the OPERA and ICARUS experiments. Since the neutrino
motion takes
place approximately on earth's surface, we deduce that the neutrinos have
a constant velocity $v= f(R_\oplus)$. Thus, if we want this not to be
superluminal but not subluminal either, at an accuracy of $10^{-7}$ of the
ICARUS result, we deduce that the parameter
$\epsilon$ in  solution subclass (\ref{solution}) must be in the interval 
$\epsilon \gtrsim 10^{-30}$ (we use relations (\ref{physicalvalues})
in order to set $c$ and $G$ to 1 and then we use the values of
$\Lambda_{eff}$, $R_\oplus$ and $M_\oplus$ in these units). Similarly,
for the Kehagias-Sfetsos solution subclass we can see that the observed
neutrino luminality is obtained for $q\lesssim-
10^{-21}$.

\section{Discussion}

In the present work we constrained  Ho\v{r}ava-Lifshitz gravity using the
data form OPERA and ICARUS neutrino speed experiments which show that
neutrinos are luminal particles, by examining the fermion
propagation in the earth's gravitational field, considering the
gravitational sector to be of Ho\v{r}ava-Lifshitz type. In particular, we
used the spherical solutions of the theory going beyond the
detailed-balance condition, and in such a background we investigated the
Dirac equation. We found that the neutrinos feel an effective metric with
respect to which they might propagate superluminally.
The reason for such a behavior is that in spherical Ho\v{r}ava-Lifshitz
solutions one obtains an extra positive term in the effective metric, and
subsequently in the fermion velocity. In general, such a result is expected
for Lifshitz-type theories and it plays the role of the ``anti-gravity''
source that is needed for superluminality, and indeed our own result in the
specific case of Ho\v{r}ava-Lifshitz gravity is in agreement with the
general qualitative result of \cite{Alexandre:2011kr,Alexandre:2011bu}.

Therefore, if one desires not to have superluminal or subluminal motion,
then the parameter $\epsilon$ in  \cite{Lu:2009em,Cai:2009pe} formulation
must be in the range $\epsilon \gtrsim 10^{-30}$, while the parameter $q$
in KS formulation of \cite{Kehagias:2009is} must be in the range 
$q\lesssim- 10^{-21}$. Clearly the excluded parameter regions are very
narrow, that is Ho\v{r}ava-Lifshitz gravity predicts luminal motion in a
large subspace of its parameter space. However, we can clearly see that
the detailed balance case (corresponding to $\epsilon=0$ in
\cite{Lu:2009em,Cai:2009pe}) is excluded (KS solution is already beyond
the detailed balance), which is in agreement with theoretical works that
exclude this case due to instabilities
\cite{Charmousis:2009tc,Bogdanos:2009uj}.

We close this work with two comments. The first is that if one desires to
apply the above analysis in neutrinos coming from
galactical distances, then he should take into account that
away from the
earth's surface the background metric is not spherical and it is not
determined by the earth anymore, but from the sun, the other planets, the
other stars etc, resulting to the Friedmann-Robertson-Walker metric, where
the above procedure results to luminal speed for massless neutrinos. This
is in agreement with anti-neutrino observations from the
SN1987A supernova, which impose the stringent constraint
$|(v-c)/c|<2\times 10^{-9}$
\cite{Longo:1987ub,Hirata:1987hu,Bionta:1987qt}.  

The second point is what version of Ho\v{r}ava-Lifshitz gravity must be
used, and which solution subclass. In the present work we
desired to provide two examples where superluminality is
theoretically possible in
Ho\v{r}ava-Lifshitz context, thus we chose a simple version of
Ho\v{r}ava-Lifshitz gravity, allowing also from a departure
from the detailed-balance condition, as a representative example, despite
the fact that more complicated extensions seem to be theoretically more
robust \cite{Blas:2009qj}. Clearly, one should repeat the above procedure
for such modified theories in order to constrain them, however the
complication of the scenario does
not allow even for an acceptable examination of general spherical
solutions.

\begin{acknowledgments}
The author wishes to thank Alex Kehagias for useful discussions and an
anonymous referee for useful comments.
\end{acknowledgments}

\end{document}